\documentclass[twocolumn]{aastex631}
\usepackage{amsmath}
\usepackage{graphics}
\usepackage{graphicx}
\usepackage{amssymb}

\begin{document}

\title{A Joint Microwave and Hard X-Ray Study Towards Understanding the Transport of Accelerated Electrons during an Eruptive Solar Flare}

\correspondingauthor{Surajit Mondal}
\email{surajit.mondal@njit.edu}

\author[0000-0002-2325-5298]{Surajit Mondal}
\affiliation{Center for Solar-Terrestrial Research, New Jersey Institute of Technology, \\
323 M L King Jr Boulevard, Newark, NJ 07102-1982, USA}

\author[0000-0002-0660-3350]{Andrea F. Battaglia}
\affiliation{University of Applied Sciences and Arts Northwestern Switzerland (FHNW), Bahnhofstrasse 6, 5210 Windisch, Switzerland}
\affiliation{Swiss Federal Institute of Technology in Zurich (ETHZ), Rämistrasse 101, 8092 Zurich, Switzerland}

\author[0000-0002-0660-3350]{Bin Chen}
\affiliation{Center for Solar-Terrestrial Research, New Jersey Institute of Technology, \\
323 M L King Jr Boulevard, Newark, NJ 07102-1982, USA}

\author[0000-0002-0660-3350]{Sijie Yu}
\affiliation{Center for Solar-Terrestrial Research, New Jersey Institute of Technology, \\
323 M L King Jr Boulevard, Newark, NJ 07102-1982, USA}

\begin{abstract}

The standard flare model, despite its success, is limited in comprehensively explaining the various processes involving nonthermal particles. One such missing ingredient is a detailed understanding of the various processes involved during the transport of accelerated electrons from their site of acceleration to different parts of the flare region. Here we use simultaneous radio and X-ray observations from the Expanded Owens Valley Solar Array (EOVSA) and Spectrometer/Telescope for Imaging X-rays (STIX) onboard the Solar Orbiter (SolO), respectively, from two distinct viewing perspectives to study the electron transport processes. 
Through detailed spectral modeling of the coronal source using radio data and footpoint sources using X-ray spectra, we compare the nonthermal electron distribution at the coronal and footpoint sources. We find that the flux of nonthermal electrons precipitated at the footpoint is an order of magnitude greater than that trapped in the looptop, consistent with earlier works which primarily used X-ray for their studies. In addition, we find that the electron spectral indices obtained from X-ray footpoints is significantly softer than the spectral hardness of the nonthermal electron distribution in the corona.
We interpret these differences based on transport effects and the difference in sensitivity of microwave and X-ray observations to different regimes of electron energies.
Such an understanding is crucial for leveraging different diagnostic methods of nonthermal electrons simultaneously to achieve a more comprehensive understanding of the electron acceleration and transport processes of solar flares. 

\end{abstract}


\section{Introduction} \label{sec:intro}

Understanding how particles are accelerated to high energies is one of the key challenges in astrophysics and space physics. The energetics of these particles and the system parameters where they are generated are difficult to reproduce through lab experiments directly. The Sun is a great laboratory to study these particle acceleration processes. Electrons and ions are accelerated to very high energies during the solar flares. There have been many studies in the past, both observational and theoretical, which have tried to understand the details of not only the acceleration process itself, but also how these particles are transported from their acceleration sites toward the solar surface and into the heliosphere. However, many questions remain. Examples include the details of trapping and scattering phenomena associated with the nonthermal electrons and how the distribution of nonthermal electrons changes due to these transport processes.

Observationally, nonthermal particles produced by flares are probed primarily using remote-sensing observations in the hard X-ray (HXR), gamma ray, and radio wavelengths, as well as \textit{in situ} measurements in the interplanetary space. At the radio wavelengths, the signatures of nonthermal electrons are attributed to either coherent emission mechanisms, including the plasma emission and the electron cyclotron maser emission, or the nonthermal gyrosynchrotron emission. The coherent emissions are generally observed at frequencies $\lesssim 2$GHz, whereas nonthermal gyrosynchrotron emission is more common at higher frequencies \citep{Bastian1998,nindos2020,gary2023}. The coherent emissions are extremely sensitive tracers of the nonthermal electrons and have been used in the past to trace electrons at or near their acceleration site(s) \citep{chen2015,chen2018} and track their transport in the corona \citep[e.g.,][]{chen2013, mccauley2017, mann2018, yu2019} and in the heliosphere \citep[e.g.][]{musset2021, badman2022, wang2023}. However, these emissions involve highly nonlinear radiation processes and are also extremely sensitive to the details of the local plasma conditions and source electron distribution; hence, it remains challenging to invert the observations and quantify the nonthermal electron distribution. Nonthermal gyrosynchrotron emission, on the other hand, is an incoherent emission mechanism and can be used to provide quantitative constraints about the nonthermal electron population \citep[see reviews by][]{Bastian1998,nindos2020,gary2023}. However, the nonthermal gyrosynchrotron emission is primarily sensitive to electrons having energies $\gtrsim100$~keV \citep{white2011,krucker2020,chen2021}. Fortunately, the HXR bands are primarily sensitive to the lower energy electrons $\lesssim100$~keV. By combining information across these two wavebands, it is possible to obtain a more comprehensive understanding of the nonthermal electron distribution over a broad energy range \citep{chen2021}.

Despite the clear synergy between the HXR and radio diagnostics, one complication is that the nonthermal HXR emission is highly weighted by the background plasma density, and is primarily dominated by emissions from the footpoints of the flare arcade when the accelerated electrons hit the dense chromosphere and loses the bulk of their energy almost instantaneously. Meanwhile, the nonthermal microwave gyrosynchrotron emission is more sensitive to the coronal magnetic field and is primarily observed in the corona. Due to this difference in their primary emission site, transport effects come into play. In some instances, particularly when the bright footprint sources are occulted, nonthermal emission from the corona can also be observed in HXRs \citep[e.g.][]{sui2003,krucker2007, liu2008, krucker2010,chen2012}. A review of nonthermal coronal HXR sources is given in \citet{krucker2008}. Studying the coronal HXR sources and quantifying their difference with respect to the footpoint HXR sources are important for investigating the transport effects. Multiple studies have reported differences between the electron spectral index inferred from the coronal and footpoint HXR \citep[e.g.][]{petrosian2002, battaglia2006}. \citet{simoes2013} and \citet{chen2012} estimated that the nonthermal electron flux in the corona is about 2--10 times larger than the nonthermal electron flux at the footpoint(s), hinting towards trapping processes in the corona. 
\citet{musset2018} used a combined radio and X-ray observation and showed that the diffusive transport model could explain the difference between the nonthermal electron numbers needed to explain both the coronal gyrosynchrotron emission and the hard X-ray footpoint and coronal sources. 
While the authors were able to explain the observations at 17 and 34 GHz reasonably well, the model and the observed spectrum showed significant differences at lower frequencies. A possible reason behind this might be the unavailability of broadband spatially resolved spectrum, due to which the authors were forced to make several assumptions, including a magnetic field model extrapolated from the photosphere. However as shown in several works like \citet{chen2020a, fleishman2020, fleishman2022}, robust estimates of the nonthermal electron distribution can be obtained using broadband imaging spectroscopy, which become a new tool to understand the transport phenomena of nonthermal electrons.

The Extended Owens Valley Solar Array (EOVSA; \citealt{Gary2018}), a solar-dedicated radio instrument, provides imaging spectroscopy observations at 451 frequencies between 1--18 GHz at a cadence of 1 s. This spectroscopic snapshot imaging capability has been shown to be revolutionary in determining spatially and temporally resolved coronal magnetic field during flares \citep{chen2020a,fleishman2020,wei2021} and properties of nonthermal electrons \citep{chen2020a,chen2020b,yu2020,fleishman2022,chen2021,lopez2022}, especially when complemented by HXR and other multi-wavelength observations. In particular, \citet{kuroda2020} used joint microwave-HXR imaging observations using data from the EOVSA and the Reuven Ramaty High Energy Spectroscopic Imager (RHESSI) to study the electron transport phenomenon from the looptop to the footpoint and found that their observations were consistent with the evolution of electrons in a simplified trap-precipitation model \citep{melrose1976}. \citet{chen2021} determined the nonthermal electron distribution in the coronal source by using a joint-fit of the microwave and HXR spatially resolved spectra. They found that while the low energy part of the distribution below approximately 100 keV is well constrained by the HXR data, the microwave observations provide the sole constraint for the electrons distribution at $>$100 keV. However both these studies focused on limb events and hence it is possible that HXR footpoints were partially occulted.

In this study, we use X-ray data from the Spectrometer/Telescope for Imaging X-rays \citep[STIX;][]{krucker2020} onboard the Solar Orbiter \citep{muller2020} and combine that with microwave data from the EOVSA. This flare event was observed close to the limb with both flare ribbons visible on the disk from the EOVSA/Earth viewpoint. From STIX's perspective, it appears as a disk event. This capability of viewing the same event from different perspectives 
allows us to provide new insights into the transport process of the accelerated electrons.

This paper is organized as follows. In Section~\ref{sec:observations}, we briefly describe the flare context, the observations, and data analysis procedures. In Section~\ref{sec:results}, we describe the results obtained from the analysis of the radio and X-ray data. In Section~\ref{sec:discussion}, we discuss the results in the context of earlier studies regarding the transport of nonthermal electrons and then conclude in Section~\ref{sec:conclusion}.

\section{Observations and Data Analysis}\label{sec:observations}

\begin{figure}
    \centering
    \includegraphics[scale=0.5]{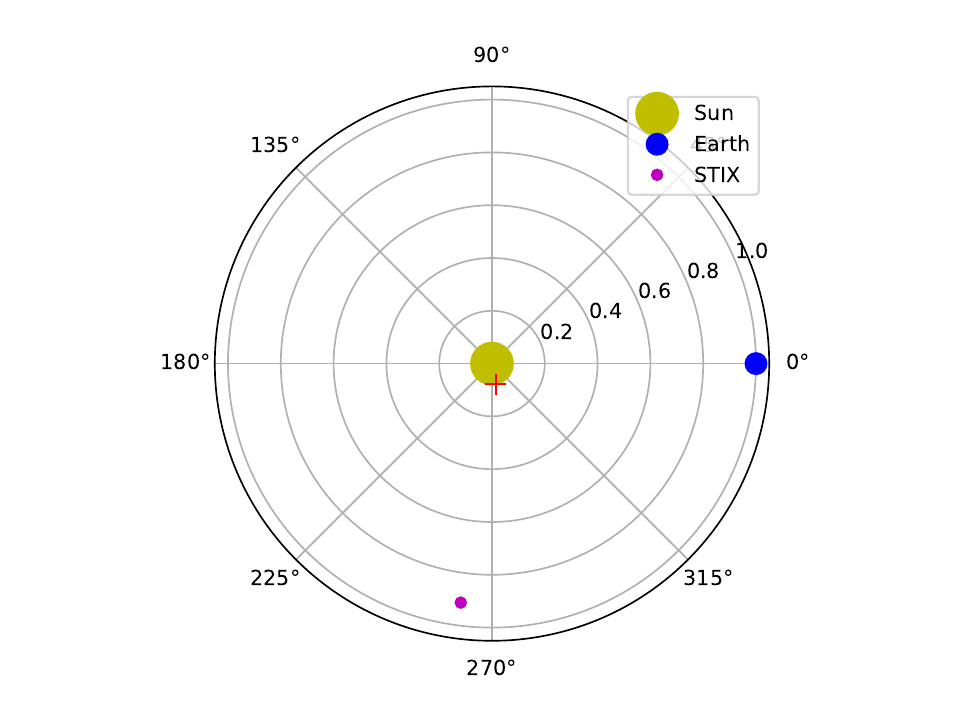}
    \caption{{ Relative location of STIX (shown with magenta), Earth (shown with blue) and Sun (shown with yellow).} The red cross indicates the flare location. For better visibility, the marker sizes used are in the same order as the true sizes, but should not be scaled.}
    \label{fig:locations}
\end{figure}

In this work we primarily use microwave data from the EOVSA, X-ray data from the STIX, extreme ultraviolet (EUV) data from the Atmospheric Imaging Assembly (AIA; \citealt{lemen2012}) onboard the Solar Dynamics Observatory (SDO; \citealt{pesnell2012}, and magnetic field measurements from the Helioseismic Magnetic Imager (HMI; \citealt{schou2012}) onboard the SDO. All data analyzed here come from 2021 May 7 between 18:45 UT and 19:05 UT. The relative position of Sun, Earth and the Solar Orbiter is shown in Figure~\ref{fig:locations}. As one can see from the figure, STIX and Earth were nearly in quadrature during this time. SDO/AIA and SDO/HMI data were downloaded from Joint Science Operations Center and calibrated using standard procedures implemented in AIAPy \citep{barnes2020,barnes2020_zenodo} and visualized with SunPy \citep{sunpy_community2020,mumford2020,mumford_2022_zenodo}. Below we describe the data reduction and analysis procedures to produce science-ready images and spectra from EOVSA and STIX, respectively.

\subsection{EOVSA Data Analysis}

\begin{figure}
    \centering
    \includegraphics[trim={1cm 0 0 0},clip,scale=0.5]{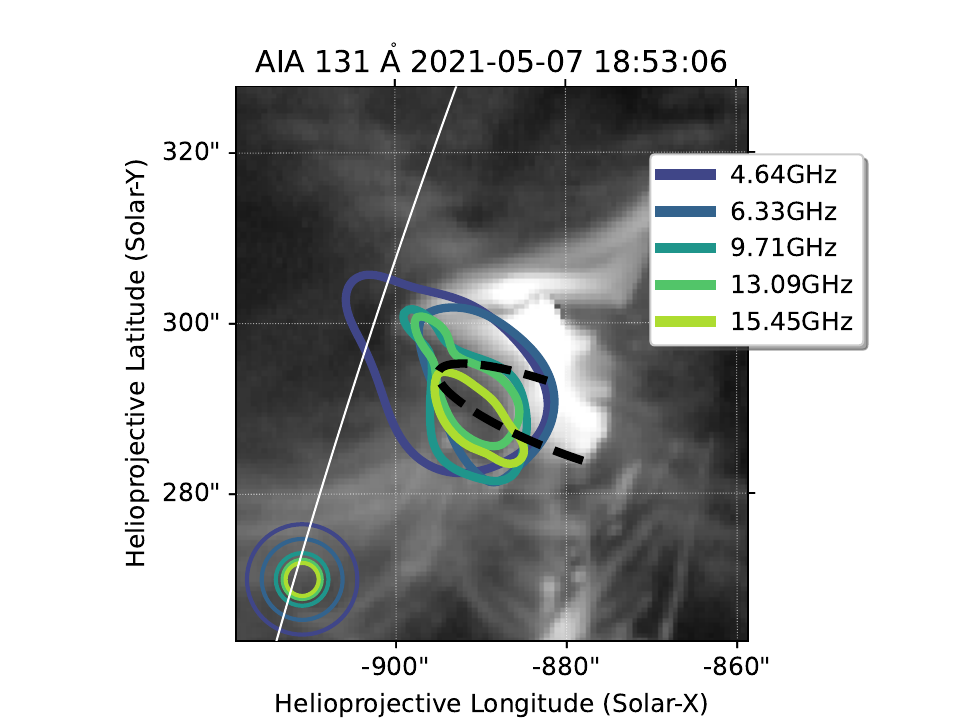}
    \includegraphics[trim={1cm 0 0 0},clip,scale=0.5]{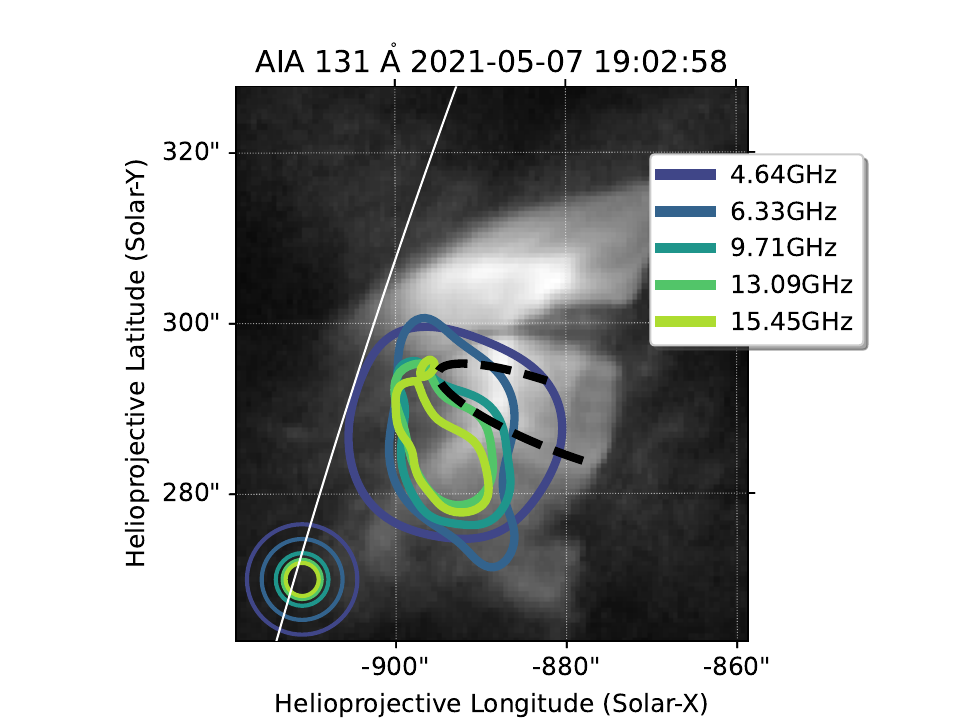}
    \caption{30\% contours of images at a few example frequencies are overlaid on an AIA 131\AA$\,$ image. The black dashed line shows the model coronal loop obtained from STIX data and shown in \citet{massa2022}. The circles shown in the bottom left corner represent the restoring beams at different frequencies. }
    \label{fig:eovsa_sample_image}
\end{figure}

Raw data at 1s resolution were obtained from the public EOVSA data archive\footnote{\url{http://ovsa.njit.edu/data.html}}. Then delay calibration, bandpass calibration and complex gain calibration was done following standard techniques\footnote{\url{http://www.ovsa.njit.edu/wiki/index.php/Calibration_Overview}}. However due to the time difference between the calibrator observations, which are used to estimate the correction terms (also known as antenna gains), and the solar observations, there can be small errors in these correction terms. These can be corrected using a procedure called self-calibration \citep{cornwell1999}. We follow the previously developed procedure for EOVSA to perform the self-calibration.\footnote{see, e.g., \url{https://github.com/suncasa/suncasa-src/blob/master/examples/eovsa_flare_slfcal_example.py}.} 
While in principle self-calibration can be done for all times, we have performed it only at a single time where the radio source was bright. The solutions are applied to the entire 20 minute time interval of interest, with the assumption that the antenna gain do not change significantly over this period (which is valid given the stability of the system and the quality of the final images). Final imaging was done at ten second cadence and ten second integration for all the frequency bands. The native resolution of the instrument at the time of the observations was $75''\times 42''$ at 1 GHz. To avoid the sense of distortion in the resulting images, all images were restored using a circular beam with a full-width-half-maximum (FWHM) size of $60''/\nu_{\rm GHz}$, where $\nu_{\rm GHz}$ is the frequency of the image in GHz. After producing these images, we performed a total power calibration. This involved adjusting the integrated flux from the image plane to match the flare's total-power flux obtained from single dish measurements. This is done for every frequency independently. In Figure~\ref{fig:eovsa_sample_image}, contours of example multi-frequency microwave images at 18:53:00 and 19:02:50 UT multiple frequencies are overlaid on a AIA 131 \AA\ image at the closest time. While images were generated for all the frequency bands available with the EOVSA, here we have chosen to show the images at only a handful of frequencies for clarity (blue to green contours in Figure~\ref{fig:eovsa_sample_image}). 

\subsection{STIX Data Analysis}
Two different data formats were used for the STIX data analysis: the compressed pixel data for producing the STIX images and the spectrogram data at 1 s resolution for the spectroscopic analysis. Because of the different heliocentric distance from the Solar Orbiter spacecraft to the Sun than the Earth, all the STIX times have been shifted by 45.8 s to compensate its shorter light travel time relative to Earth.

The STIX images have been reconstructed using the CLEAN algorithm \citep{hogbom1974}, with a beam FWHM width of 14.6 arcsec, which corresponds to the resolution of the subcollimator 3 \citep{krucker2020}. The finest two subcollimators, labeled 1 and 2, are not yet fully calibrated and have been excluded for the imaging reconstruction. The STIX observations used in this paper has been recorded during the Solar Orbiter’s cruise phase, which is outside the nominal Solar Orbiter science window. During the flare, the spacecraft was at a heliocentric distance of 0.91 AU from the Sun. At such a large distance, the STIX Aspect System \citep[SAS;][]{warmuth2020} was not fully functional. Therefore, the STIX images have been manually shifted{\footnote{manual shift of 20 arcsec East and 14 arcsec North was needed for alignment}} and co-aligned to the reprojected AIA 1600 \AA{} map closest in time to the STIX nonthermal peak \citep[more details can be found in][]{battaglia2021,massa2022}. 

The standard software for solar spectral analysis, Object Spectral Executive, \texttt{OSPEX} \citep{tolbert2020} was used for the spectral modeling of the STIX observations. Because of the presence of yet unknown systematic effects in the calibration, we assumed the existence of an additional 5\% source of systematic error that we added in quadrature to the errors from the photon counting statistics.

\section{Results} \label{sec:results}

\subsection{Microwave, X-ray and EUV lightcurves}

\begin{figure*}
    \centering
    \includegraphics[scale=0.6]{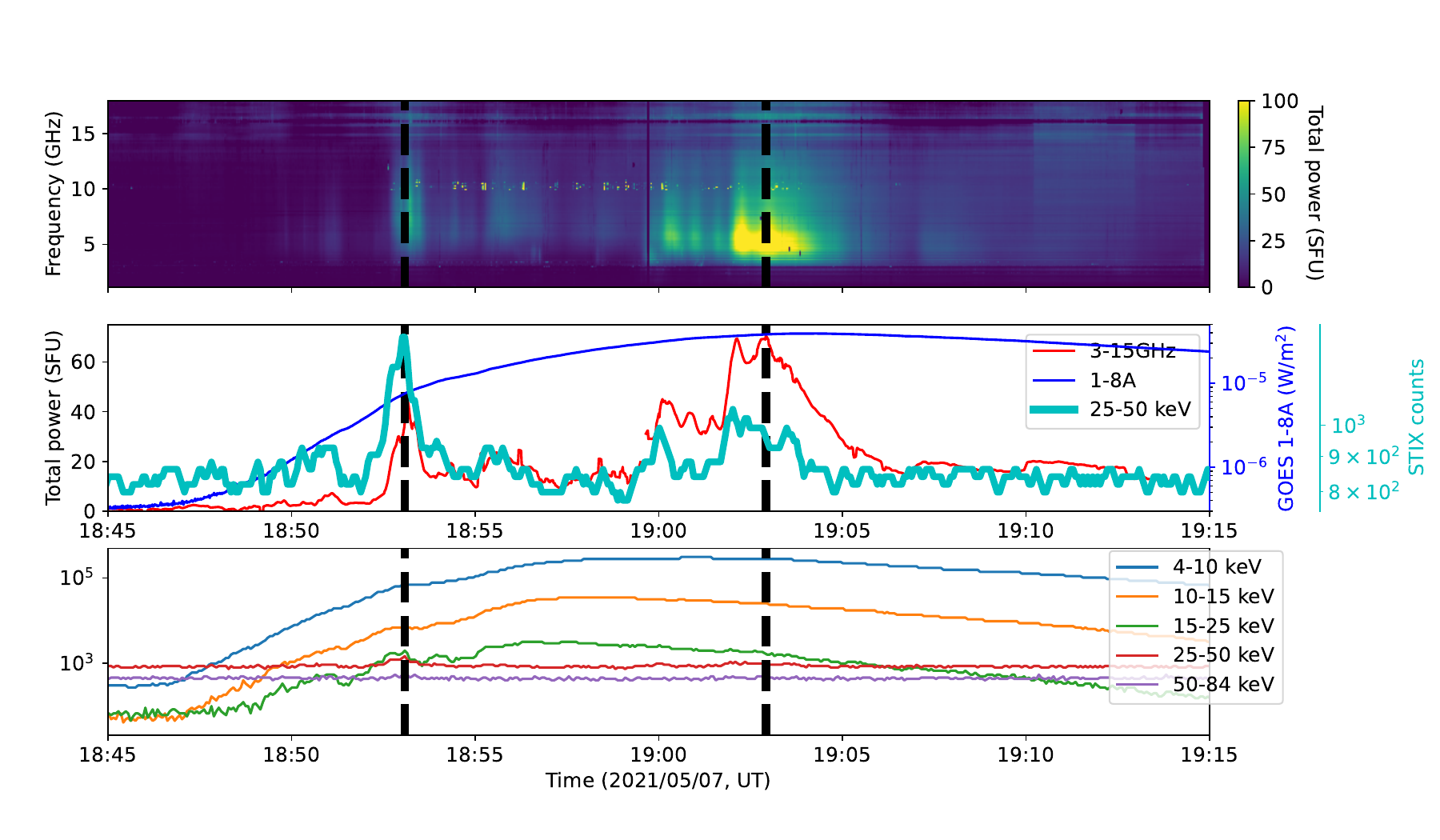}
    \caption{Top panel: Dynamic spectrum obtained from EOVSA. The colorbar has been saturated at 100 SFU. Middle panel: Frequency averaged EOVSA lightcurve is shown with red. GOES X-ray lightcurve is shown with blue. STIX lightcurve at 25--50 keV is shown with a {thick cyan line}. Bottom panel: STIX lightcurves at different energy bands are shown. The black dashed lines show the times studied here.}
    \label{fig:eovsa_ds}
\end{figure*}

 In Figure~\ref{fig:eovsa_ds}, we show the total power dynamic spectrum from the EOVSA in the upper panel. The frequency averaged microwave lightcurve and 1--8 \AA\ soft X-ray (SXR) lightcurve obtained from the GOES satellite are shown in the middle panel in red and blue colors, respectively. With { a thick cyan line} we have shown the STIX lightcurve at 25--50 keV as well. In the bottom panel we show the lightcurves obtained from STIX at different X-ray energy bands from 4--10 keV to 50--84 keV. The black dashed lines show the times studied here. 

\begin{figure*}
    \centering
    \includegraphics[scale=0.5]{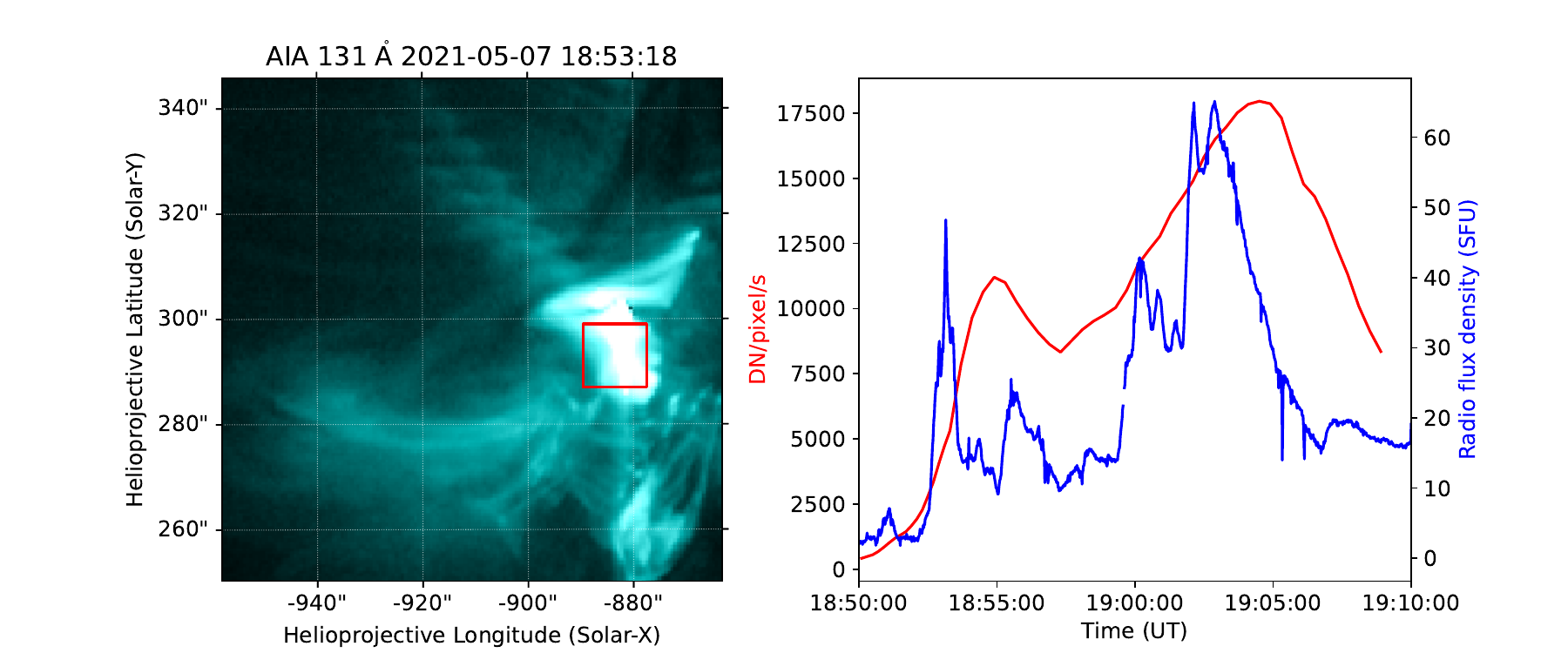}
    \caption{Left panel: AIA 131\AA$\,$ image. The {red} box shows the region from where the lightcurve has been extracted. Right panel: 131\AA$\,$ lightcurve extracted from the {red box is shown with red}. The {blue} line shows the frequency averaged radio lightcurve.} 
    \label{fig:aia_lightcurve}
\end{figure*}

The GOES 1--8 \AA\ lightcurve peaks around 19:03 UT. The 4--10 keV X-ray lightcurve obtained by STIX also peaks at a slightly earlier time. The higher-energy 10--15 and 15--25 keV lightcurve shows a peak around 18:58:30 UT and 18:57:30 UT, respectively. The trend of an earlier peak time for a higher X-ray energy band has been commonly observed in many other flares \citep[e.g.][]{neupert1968,dennis1993}. These energy bands also show a smaller peak at 18:53:00 UT, around the time when a prominent microwave burst is seen with the EOVSA. Interestingly, this microwave peak also coincides with an enhancement in the 25--50 and 50--84 keV HXR lightcurves. From the middle panel of Figure \ref{fig:eovsa_ds}, it is clear that the microwave lightcurve has striking similarities with the 25-50 keV HXR lightcurve from STIX. Such similarities in microwave and HXR lightcurves during the flare impulsive phase have been reported previously, which were suggested as the signature of emissions by the same population of flare-accelerated electrons \citep[e.g.,][]{minoshima2008,liu2015,Gary2018}.



In the right panel of Figure~\ref{fig:aia_lightcurve}, we show the lightcurve at 131\AA$\,$ in red. The lightcurve is extracted from the region inside the box shown in the left panel. This box is co-spatial with the microwave emission source observed by EOVSA.  The frequency-averaged 3--15 GHz EOVSA lightcurve is shown in red. It is interesting to note that the EUV lightcurve also shows two peaks similar to the two major peaks seen in the radio and HXR lightcurves. However the EUV peaks are much broader and smoother and come at a slightly later time than the radio and X-ray peaks. It is possible that this is because both the radio and STIX lightcurves are due to nonthermal emission (will be discussed later in the text) from energetic electrons that vary at short scales, while the AIA 131 \AA\ light curve reflects the temporal variation of the hot, $\sim$10 MK plasma accumulated in the flare arcade through chromospheric evaporation \citep[see, e.g.,][for a review]{benz2017}. 

\begin{figure*}
    \centering
    \includegraphics[scale=0.5]{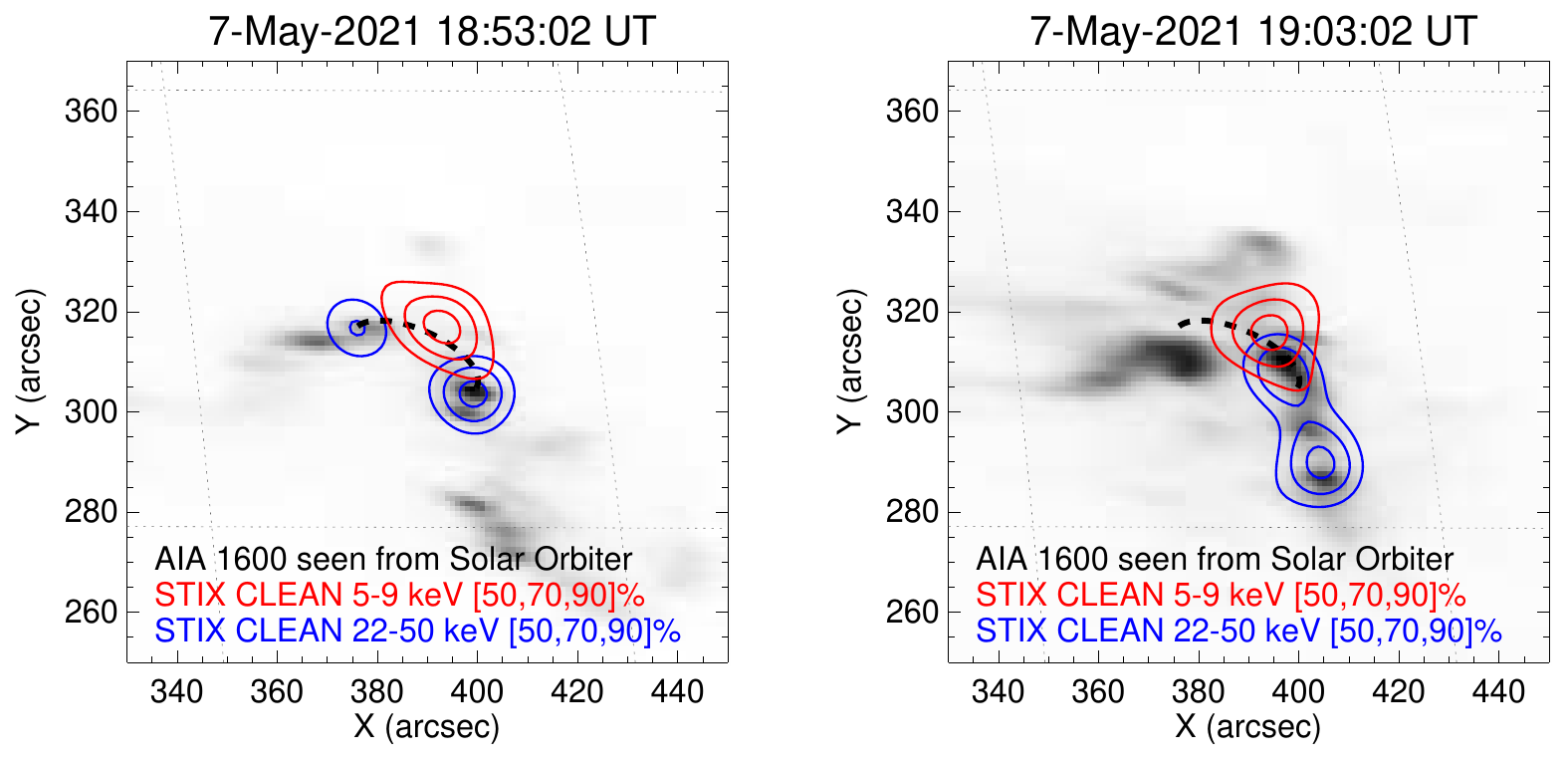}
    \caption{Solar Orbiter/STIX reconstructed images, as 50, 70 and 90\% contours of the maximum, on top of the reprojected SDO/AIA 1600 \AA{}. The two panels show the reconstructed images around two different instances: 18:53:02 (\emph{left}) and 19:03:02 (\emph{right}), which correspond to the times used for modeling the flux density. The red and blue contours show the images reconstructed within the energy range from 5 to 9 keV (thermal emission) and from 22 to 50 keV (nonthermal emission), respectively.}
    \label{fig:stix_images}
\end{figure*}

\subsection{Microwave, X-ray and EUV images} \label{subsec:images}

In Figure~\ref{fig:eovsa_sample_image} we have overlaid contours of microwave images at five representative frequencies on top of an AIA 131\AA$\,$ image at the closest time. We also overlaid the coronal loop estimated in \citet{massa2022}. The model coronal loop was generated using the STIX data under the assumption of a semi-circular loop connecting the HXR footpoints. While the loop was constructed only for 18:51 UT, we have plotted the same loop at both the two times studied here---18:53 UT and 19:03 UT---for reference. At 18:53:00 UT, we find that the microwave source is most likely a looptop source, assuming that the estimated semi-circular loop corresponds to the ``true" coronal loop during this time. Due to projection effects and limited instrumental resolution, however, it is unclear whether the source is located at or above the looptop.

In Figure~\ref{fig:stix_images}, we show the STIX images, as red and blue contours, on top of the re-projected AIA 1600 \AA{} maps in the Solar Orbiter's view. In order to have sufficient counting statistics for the image reconstruction, we integrated 
over one minute around the central times used for modeling the X-ray spectra, which are 18:53:02 UT and 19:03:02 UT, respectively. The geometry highlighted in the left panel is consistent with the standard flare picture: Two footpoints (blue contours) seen in the 22--50 keV HXR band that are co-spatial with the UV ribbons shown in the AIA 1600 \AA{} map. They represent the anchoring points of the newly formed flare arcade visible in the 5--9 keV SXR band (red contours). It is interesting to note that the western footpoint appears brighter than the eastern one. For the later time at 19:03:02 UT (right panel), instead, the eastern footpoint is not visible anymore. However, due to the limited dynamic range of the instrument, the absence of the eastern footpoint does not necessarily mean that there is no emission at all, but perhaps much weaker. 
This interpretation is consistent with the location of the flare loop shown by the red contours, which did not change significantly relative to the previous time frame. Nevertheless, the appearance of another HXR footpoint centering at [410$''$, 285$''$] in STIX's perspective is indicative of a change of connectivity of the post-flare arcade. This is also supported by the fact that at 19:02:50 UT, we see that the radio source has shifted by a few arcseconds towards the south from its position at 18:53:00 UT (Figure~\ref{fig:eovsa_sample_image}). This suggests that the same coronal loop model shown in Figure~\ref{fig:eovsa_sample_image} derived from the earlier time of 18:53:00 UT may not be valid at this time.


\subsection{Spectral modeling of microwave data} \label{subsec:fitting}

We have performed spectral modeling of the microwave data during the two largest peaks seen in the radio lightcurve, during 18:53 and 19:03 UT, focusing on the loop-top source. Fast codes developed by \citet{fleishman2010} and \citet{kuznetsov2021} were used for calculating the gyrosynchrotron emission. We assume that the nonthermal electron distribution is isotropic and has a power-law form $f(E) = dn_{\rm nth}/dE \propto E^{-\delta'}$, where $n_{\rm nth}=\int_{E_{\rm min}}^{E_{\rm max}}f(E)dE$ is the total nonthermal electron density, with ${E_{\rm min}}$ and ${E_{\rm max}}$ being the low- and high-energy cutoff, respectively, and $\delta'$ is the power-law index. ${E_{\rm min}}$ and $E_{\rm max}$ are fixed at 10 keV and 10 MeV respectively. The depth along the line of sight (LOS) and the temperature is fixed at 7.2 Mm and 5 MK respectively. The free parameters used for the fitting are the magnetic field strength ($B$), density of thermal electrons ($n_{th}$), total density of nonthermal electrons ($n_{\rm nth}$), power-law index ($\delta'$), and angle between the LOS and the magnetic field ($\theta$). We also assume that the source along the LOS is homogeneous. 

\begin{figure*}
    \centering
    \includegraphics[width=1\textwidth]{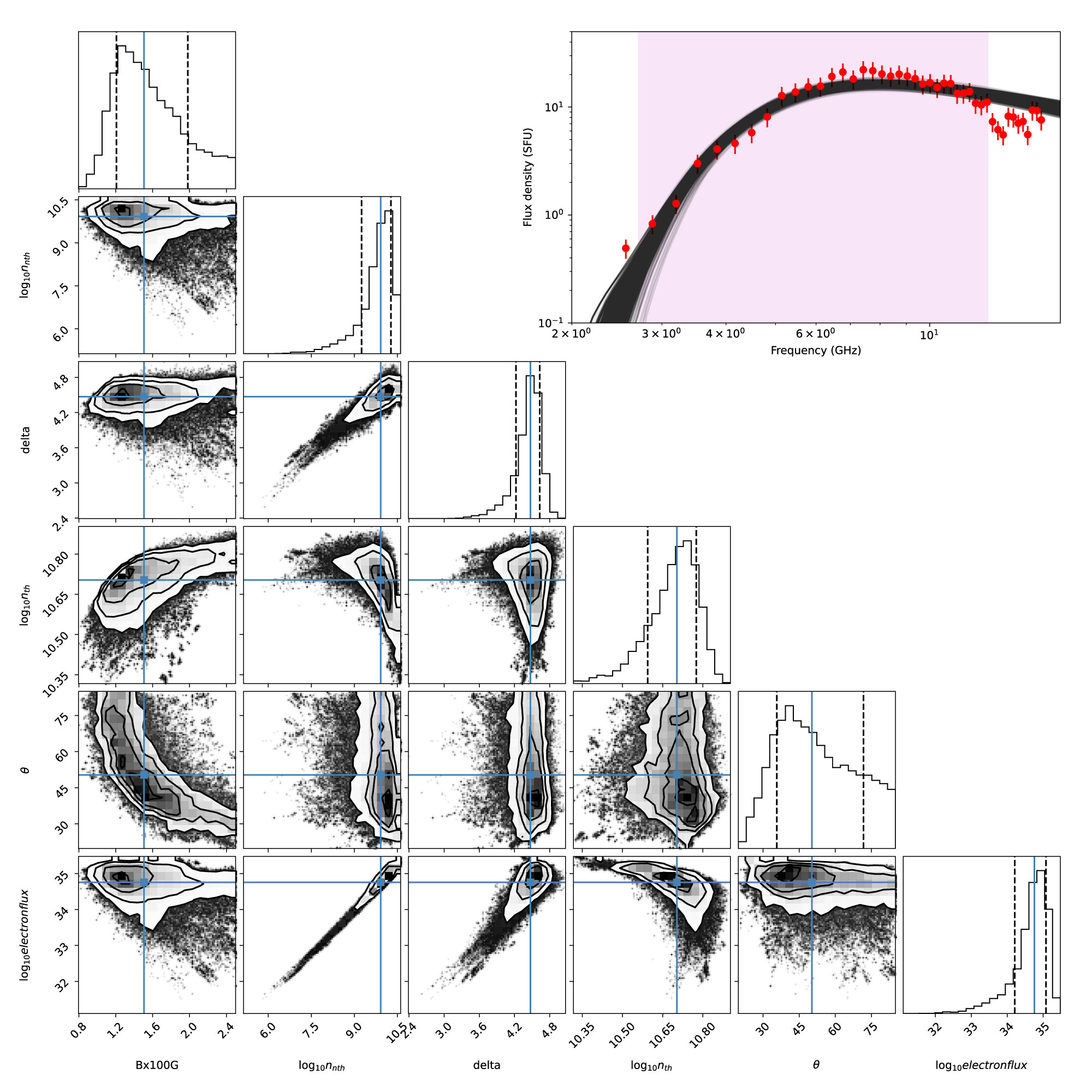}    
    \caption{In the left we show the joint and marginalized probability distribution for different fitted parameters corresponding to the loop top source at 18:53:00--18:53:10. The top right panel shows the observed spectrum (red) and the modelled spectra (black) corresponding to different parameter combinations obtained from the MCMC analysis.}
    \label{fig:param_185300}
\end{figure*}

\begin{figure*}
    \centering
    \includegraphics[width=1\textwidth]{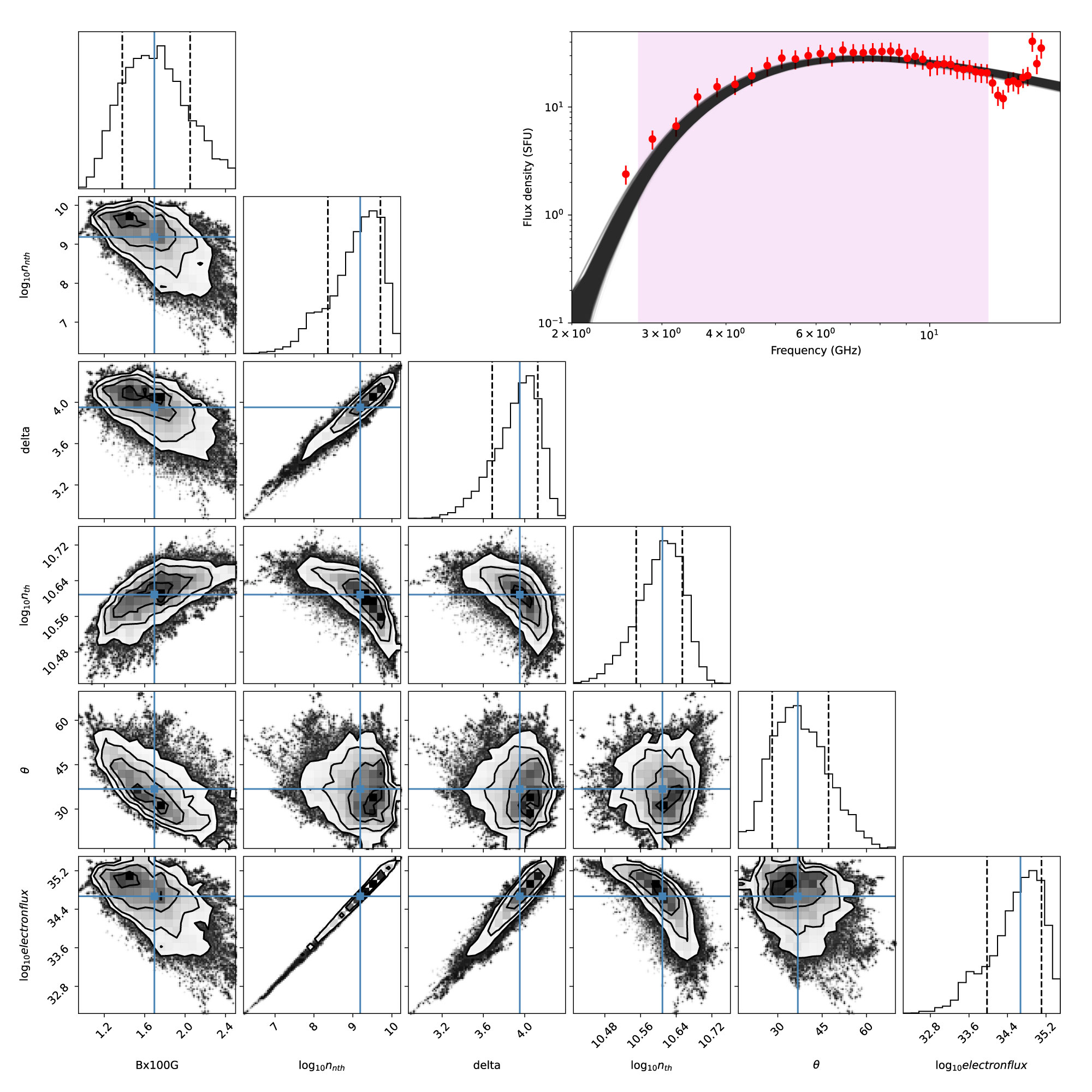}
    \caption{Results shown for the loop top source at 19:02:50--19:03:00 in the same format as that in Fig. \ref{fig:param_185300}.}
    \label{fig:param_190250}
\end{figure*}

In this work, we choose to model the flux density inside a circle centered at (-893$''$, 288$''$) with radius of 10$''$ for two times 18:53:00--18:53:10 UT and 19:02:50--19:03:00 UT. The circle is centered around the image peak at $\sim6$ GHz, the peak frequency seen in the EOVSA dynamic spectrum. The two times selected for spectral modeling are during the peak of the frequency-averaged EOVSA lightcurve in 3--15 GHz. Following \citet{chen2020a}, we perform a Markov chain Monte Carlo (MCMC) analysis to explore the parameter space. In Figures \ref{fig:param_185300} and \ref{fig:param_190250}, we show the posterior probability distribution of the free parameters and 
their marginalized probability distribution. The observed spectrum and the various model spectra using each ordered parameter set obtained from the MCMC analysis are shown in red and black respectively. The error bars have been derived by adding the r.m.s. noise level in the image and an assumed 20\% systematic error of the absolute brightness temperature in quadrature. We restrict our fitting in the frequency range inside the magenta box shown in Figures~\ref{fig:param_185300} and \ref{fig:param_190250}. The primary reason for excluding the high frequency part above 15 GHz is because, at these frequencies, significant artifacts due to increased system noise and instability are present, which can be seen from the EOVSA dynamic spectrum given in the top panel of Figure~\ref{fig:eovsa_ds}. The lower frequencies are also affected by the presence of significant radio frequency interference (RFI) noise and hence are not been used in this analysis. The fitted parameters for these two times are given in Table~\ref{tab:model_parameters}.


\begin{table*}[]
    \centering
    \begin{tabular}{c|c|c|c|c|c|c}
    \hline
         Time & $B$(G) & $\log_{10}n_{nth}$(cm$^{-3}$) & $\delta^{'}$ & $\log_{10}n_{th}$(cm$^{-3}$) & $\theta$(deg) & $\log_{10}$electron flux  \\ \hline
         18:53:00--18:53:10 & $150^{+50}_{-31}$ &  $9.9^{+0.36}_{-0.7}$ & $4.47^{+0.16}_{-0.25}$ & $10.70^{+0.07}_{-0.11}$ & $50^{+22}_{-15}$ & $34.76^{+0.33}_{-0.58}$ \\
         19:02:50--19:03:00 & $169^{+37}_{-33}$ &  $9.2^{+0.5}_{-0.8}$ & $3.95^{+0.18}_{-0.28}$ & $10.60^{+0.04}_{-0.06}$ & $36^{+10}_{-9}$ & $34.6^{+0.4}_{-0.7}$\\
         \hline
    \end{tabular}
    \caption{The model parameters and their associated uncertainties are given. The total parameter range including the uncertainties correspond to the 70\% confidence interval.}
    \label{tab:model_parameters}
\end{table*}

\subsection{X-ray spectral analysis}

\begin{figure*}
    \centering
    \includegraphics[scale=0.95]{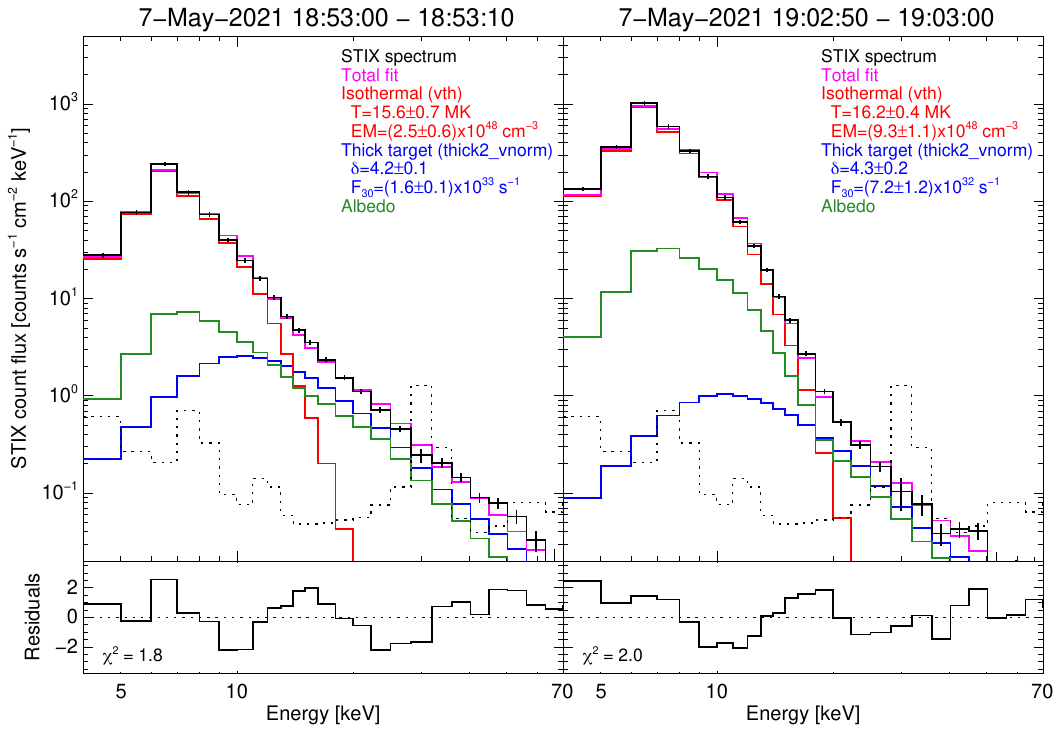}
    \caption{STIX X-ray background-subtracted spectra (solid black) for the two times considered in this paper. The low energy part of the spectra is fitted with an isothermal model (red), whereas the high energy part with a cold thick target model (blue). During both intervals, we additionally included the albedo component (green). The dotted black curve represents the background that has been subtracted from the data. Below each plot we report the residuals, observations minus total fit, in units of the standard deviation calculated from the counting statistics. The resulting fit parameters are shown in the legends.}
    \label{fig:stix_spectra}
\end{figure*}

The X-ray spectral analysis results are represented in Figure~\ref{fig:stix_spectra}. The accumulation times used in this case are exactly the same as those used for the EOVSA radio spectra. At low energies, both spectra are well fitted with an isothermal component, for which we assumed coronal abundances from the CHIANTI database (10.0.1) \citep{dere1997,zanna2020}. According to our interpretation of the observed HXR sources as footpoint emission, we fit the high-energy part of the spectrum with the cold thick-target model. This nonthermal interpretation is also consistent with the impulsive time evolution of the HXR and radio light curves. { While we have used a single isothermal component to fit the data of the second time interval for consistency, we find that addition of a second isothermal component reduces the $\chi^2$ from 2 to 0.7, about a third of that obtained with a single isothermal component. We have provided the results obtained with two isothermal components in the Appendix. However, we note here that the main results of this work do not change with the addition of the second isothermal component.}


\section{Discussion}\label{sec:discussion}

Using the spectral analysis results of the microwave looptop source, we estimate the total number of electrons that reach the footpoints under the assumption of free-streaming and equipartition. {Equipartition in this context refers to the fact that we assume that the fraction of electrons moving along the three orthogonal directions at the point of injection is equal. In other words, one-third of the total electron flux at the coronal source reaches the two footpoints.} First, we estimate the source area $A$ using the FWHM size of the radio source at 13 GHz, where the resolution is sufficient to resolve the source well. The FWHM of the deconvolved source at 18:53:00 and 19:02:50 are $7''.9 \times 3''.4$ and $10''.6 \times 6''.0$ respectively. Then we calculate the integrated electron flux above a selected energy $E_0$, {  represented by $\mathrm{F_{E_0}}$}, streaming down to the footpoints for each of the solutions obtained from the MCMC analysis discussed in Section \ref{subsec:fitting}, using the formula given by

\begin{equation}
\begin{aligned}
    \mathrm{F_{E_0}}=& { A}\int_{E_0}^{E_{max}}v f(E) dE \\ & =\frac{{ A}}{3} n_{nth} \frac{\delta'-1}{\delta'-3/2}\sqrt{\frac{2}{m}}\left(\frac{E_{min}}{E_0}\right)^{\delta'-1} \\
    &\left(\frac{E_0}{1keV}\times1.602 \times 10^{-19}\times10^7\times 10^3\right)^{1/2}.
    \end{aligned}
\label{eq:flux}
\end{equation}

{  In Equation \ref{eq:flux}, $v$ and $m$ are used to represent the velocity and mass of electrons. All other quantities are same as that described in Section \ref{subsec:fitting}. $n_{nth}$ and $m$ are in CGS units, whereas $E_{min}$, $E_0$ are in units of $keV$ and $\mathrm{F_{E_0}}$ has units of $s^{-1}$.}
The posterior distribution of the integrated nonthermal electron flux above 30 keV reaching the footpoints as predicted from the spectral modeling results of the looptop microwave source are shown in the bottom right panels of Figures \ref{fig:param_185300} and \ref{fig:param_190250}. In Table \ref{tab:electron_flux}, we provide the predicted electron flux and observed electron flux above 30 keV at the footpoints for ease of the reader. The 30 keV threshold for calculating the nonthermal electron flux is chosen based on the HXR spectra to ensure that there is no contamination from the thermal electrons in the HXR-calculated electron flux. We obtain the ratio of the electron flux predicted using the spectral modeling results of the looptop microwave source and that obtained by modeling the HXR footpoint sources. The ratio (henceforth referred as $R$) estimated for the two selected intervals 18:53:00--18:53:10 UT and 19:02:50--19:03:00 UT are {$33^{+43}_{-25}$ and $55^{+111}_{-46}$}. The uncertainties correspond to a 70\% confidence interval. The large uncertainties indicate that there are significant degeneracies in the multi-parameter space, which can also be seen in Figures \ref{fig:param_185300} and \ref{fig:param_190250}. Nevertheless, we find that $R$ is consistently greater than 1 and is consistent with previous analysis based on both coronal and footpoints HXR sources \citep{simoes2013,chen2012}.

\begin{table*}[ht]
    \centering
    \begin{tabular}{c|c|c}
        & 18:53:00--18:53:10 & 19:02:50--19:03:00  \\ \hline \hline
        $\log_{10}$predicted electron flux ($>$30 keV)  & $34.76^{+0.33}_{-0.58}$ & $34.6^{+0.4}_{-0.7}$\\ 
        $\log_{10}$observed electron flux ($>$30 keV) & $33.20^{+0.03}_{-0.03}$ & $32.8^{0.1}_{-0.1}$\\
    \end{tabular}
    \caption{First row: Nonthermal electron flux (above 30 keV) arriving at the footpoints predicted using the microwave looptop source for the two main peaks. Second row: Values constrained from the observed footpoint HXR sources.}
    \label{tab:electron_flux}
\end{table*}

%


One may attribute this difference in electron flux to transport effects from the looptop to the footpoints that depart from our assumption of equipartition and free-streaming while predicting the electron flux arriving at the footpoints. 
Alternatively, the difference between the predicted and observed nonthermal electron flux at the footpoints may be explained based on the fact that the gyrosynchrotron emission is primarily sensitive to mildly relativistic electrons (energies $\gtrsim 100$ keV), whereas the observed HXR flux in $\sim$10--70 keV are mainly contributed by non-relativistic electrons in the deca-keV range. In fact, as shown in \citet{chen2021}, that for modeling the observed HXR and microwave spectrum simultaneously of a coronal source, a broken powerlaw model of the nonthermal electron distribution is required. Consequently, the presumption that the nonthermal electron distribution conforms to a single power-law may be overly simplistic and could partly account for the discrepancies observed in the nonthermal flux as derived from microwave and hard X-ray (HXR) observations. 
 
Turning to transport effects, \citet{musset2018} showed that in the strong diffusion regime by turbulent pitch angle scattering, higher energy electrons are more efficiently trapped compared to lower energy ones. This is because the pitch angle diffusivity is proportional to $\beta^{-3}\Gamma^{-2}$ \citep{minoshima2008}, where $\beta=v/c$ is the ratio between the velocity of the electron and the speed of light in vacuum and $\Gamma=(1-\beta^2)^{-1/2}$ is the Lorentz factor. This implies that an anisotropic distribution of low-energy electrons quickly becomes isotropic compared to their high-energy counterparts. Electrons with small pitch angles will be quickly precipitated, leading to further diffusion of electrons into low-pitch angles. This process leads to an excess of high-energy electrons trapped in the coronal source. Due to this reason, the precipitation rate is also a function of energy and is proportional to $E^{-3/2}$, leading to an excessive precipitation of lower energy electrons at the footpoints that produce thick-target HXR emission compared to their higher energy counterpart. 

Here we also find, both from the radio and X-ray data, that nonthermal electron distribution seems to be harder at 19:03:00 compared to that at 18:53:00, although strictly speaking the spectral hardness is consistent between two times considering the large uncertainties. We argue that this spectral hardening can also be reconciled with the difference between $R$ estimated at the two times based on the dependence of trapping efficiency and particle energy as well. A harder nonthermal electron distribution implies a relatively smaller number of lower energy electrons compared to the higher energy ones. Also, it is the higher energy electrons that are trapped better \citep{lee2002,white2011,nindos2020}. Since no such transport effects were considered during our calculation of predicted nonthermal electrons arriving at footpoint, the $R$ we obtained at 19:03 UT may be over-estimated. A corollary of this hypothesis is that if we consider the flux of electrons in a lower energy range in, say, between 20--30 keV, then the value of R could be smaller than that obtained when we consider electrons with energies above 30 keV. We were only able to verify this corollary in the case of 18:53:00--18:53:10, as the uncertainties in the estimated nonthermal electron flux were too large at 19:02:50--19:03:00 for the HXR data. We find that the median ratio at 18:53:00--18:53:10 for electrons having energies between 20--30 keV is approximately 18. This is smaller than the median $R$ found earlier considering electrons above 30 keV and is consistent with our hypothesis presented earlier.


The increased trapping efficiency of higher energy electrons can also explain the difference between the power-law index estimated from the microwave looptop source (denoted as $\delta_{\mathrm{radio}}^{'\rm LT}$) and that estimated from the X-ray footpoint source (denoted as $\delta_{\mathrm{xray}}^{\rm FP}$). The different notations are used because the former corresponds to the differential electron density distribution ($dn_{\rm nth}/dE$), while the latter is for the differential electron flux distribution ($vdn_{\rm nth}/dE$). If the same nonthermal electron distribution responsible for the looptop microwave emission produces the X-ray emission at the footpoints, the predicted spectral index of the electron flux distribution at the footpoint would be $\delta^{\rm FP}_\mathrm{predict}\approx\delta^{'\rm LT}_\mathrm{radio}-0.5$ \citep{oka2018}. This translates to $\delta^{\rm FP}_\mathrm{predict}=3.9\pm 0.2$ and $3.4\pm 0.2$ at 18:53 and 19:03 UT respectively.  
Compared to the spectral indices derived from the observed X-ray footpoint sources $\delta_\mathrm{xray}^{\rm FP}=4.2\pm0.1$ and $4.3\pm0.2$ for the two intervals, respectively, those estimated using the microwave looptop source seems to be a bit harder, although strictly speaking they are consistent within uncertainties at the interval 18:53:00--18:53:10
One of the explanations behind this difference is that the low energy electrons can more easily escape the looptop regions and precipitate at the footpoints, leading to a softer spectrum at the footpoint. Similar results were also obtained on \citet{kuznetsov2015} and \citet{musset2018}. {And this is consistent with the fact that for the interval 19:02:50--19:03:00, we find that the nonthermal electron distribution estimated from radio data is considerably harder than that estimated for the interval 18:53:00-18:53:10.}

\section{Conclusion} \label{sec:conclusion}

In this study, we have used simultaneous imaging and spectroscopy in the microwave and X-ray wavelengths to study the flare transport processes. We analyzed the microwave spectra obtained by EOVSA to determine the magnetic field and the nonthermal electron distribution of the looptop microwave source. At X-ray wavelengths, STIX imaging shows that the HXR source is dominated by footpoint sources. Hence, the HXR spectrum is modeled under the thick-target bremsstrahlung assumption, which gives the nonthermal electron distribution precipitated to the chromosphere. The resulting footpoint nonthermal electron distribution constrained by X-ray data is then compared to those predicted by the looptop microwave source under the assumption of free-streaming and equipartition. We find that the predicted electron flux reaching the footpoint using the looptop source is an order of magnitude higher than that obtained from the observed X-ray footpoint source. This is interpreted as a result of transport effects. 
This result, once again highlights the importance of taking into account transport effects while comparing nonthermal electron distributions across different spatial locations and also across wavebands which are sensitive to different energy ranges.

\begin{acknowledgments}
The Expanded Owens Valley Solar Array (EOVSA) was designed, built, and is now operated by the New Jersey Institute of Technology as a community facility. EOVSA Operations are supported by NSF grant AGS-2130832 and NASA grant 80NSSC20K0026 to NJIT. Solar Orbiter is a space mission of international collaboration between ESA and NASA, operated by ESA. The STIX instrument is an international collaboration between Switzerland, Poland, France, Czech Republic, Germany, Austria, Ireland, and Italy.  S.M., B.C., and S.Y. are supported by NASA grant 80NSSC20K1318 and NSF grant AST-2108853 awarded to NJIT. AFB is supported by the Swiss National Science Foundation Grant 200021L\_189180 for STIX. CHIANTI is a collaborative project involving George Mason University, the University of Michigan (USA), University of Cambridge (UK) and NASA Goddard Space Flight Center (USA).
This research used version 4.0.5 \citep{mumford_2022_zenodo} of the SunPy open source software package \citep{sunpy_community2020}.
This research used version 0.6.4 \citep{barnes2020b} of the aiapy open source software package \citep{barnes2020}.
This research has made use of NASA's Astrophysics Data System Bibliographic Services.
\end{acknowledgments}

\software{Numpy \citep[][]{Harris2020},  
          Scipy \citep[][]{Scipy2020},
           Python 3 \citep[][]{python3},
           Matplotlib \citep[][]{Hunter:2007},
           Sunpy \citep[][]{sunpy_community2020},
           Astropy\citep{astropy2013,astropy2018,astropy2022},
           Aiapy\citep{barnes2020b},
           Corner\citep{corner},
           Lmfit \citep{lmfit2023},
           Emcee \citep{emcee2013}
           }

\bibliography{sample631}{}
\bibliographystyle{aasjournal}


\appendix
{
\section{Results obtained using two isothermal components for fitting STIX data at 19:02:50--19:03:00}

\begin{figure*}
    \centering
    \includegraphics[scale=0.95]{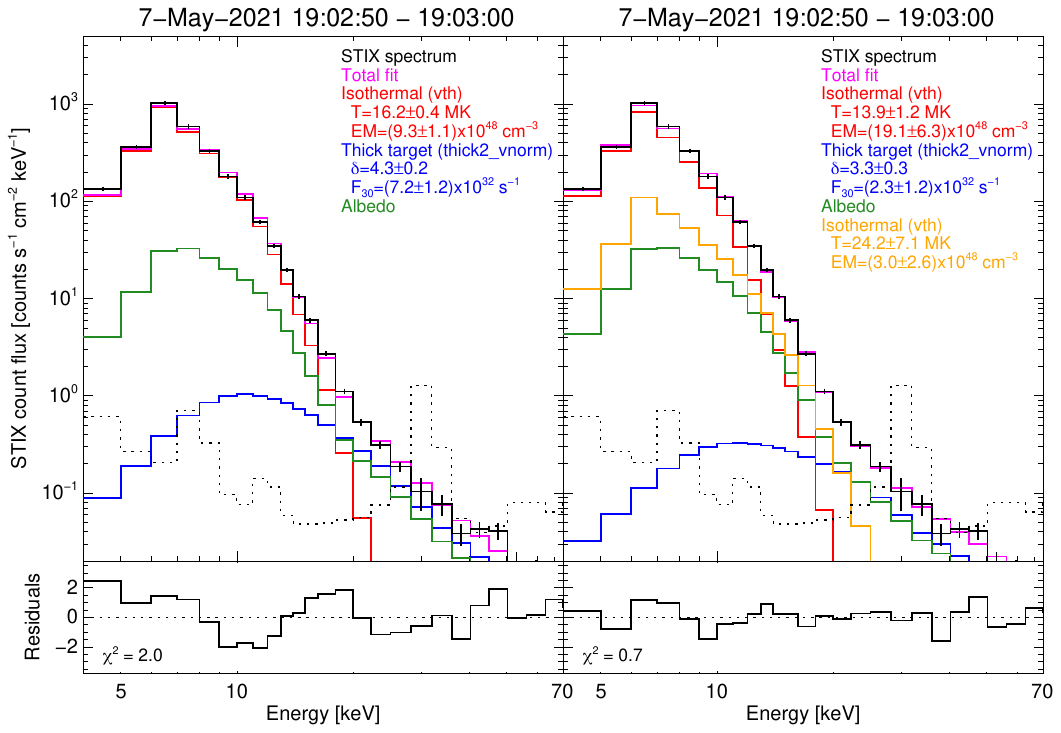}
    \caption{STIX X-ray background-subtracted spectra (solid black) for 19:02:50--19:03:00 is shown. The format of the figure is same as that in Figure \ref{fig:stix_spectra}. The left and right panels correspond to the fits made using a powerlaw distribution of nonthermal electrons and, one and two isothermal components respectively.}
    \label{fig:stix_spectra_app}
\end{figure*}

In the main text, we use one isothermal component for fitting the STIX data at 19:02:50--19:03:00. The results are shown in the right panel of Figure~\ref{fig:stix_spectra}. For completeness, in the right panel of Figure \ref{fig:stix_spectra_app}, we also show the results obtained when we use two isothermal components for fitting the same spectrum. For ease of comparison, we have also provided the data and fit results using one isothermal component on the left panel of the same figure. As mentioned earlier, we find that the addition of the second isothermal component reduces the $\chi^2$ significantly. We also find that while the best-fit $\delta$ is much higher for the two isothermal case, the total nonthermal electron flux is about three times smaller than the single isothermal component case and poorly constrained. However, even if we use these numbers, we find that $R=93^{+475}_{-82}$, which is also consistent with the key point of this work that the nonthermal electron flux estimated using microwave data from the coronal cycle is consistently higher than that estimated using X-ray data at the footpoint.}

\end{document}